\input{aipcheck} 
 
\documentclass[ 
    ,final            
  ] 
  {aipproc} 
 
\layoutstyle{6x9} 
 
\begin{document} 
 
\title{Neutrino emissivity and bulk viscosity of \\ 
iso-CSL quark matter in neutron stars} 
 
\classification{04.40.Dg, 12.38.-t, 26.60.+c} 
\keywords{neutrino cooling, bulk viscosity, dense quark matter,  
color superconductivity} 
 
\author{David B. Blaschke}{ 
  address={Institute for Theoretical Physics, University of Wroclaw,  
50-204 Wroclaw, Poland\\ 
Bogoliubov Laboratory for Theoretical Physics, JINR Dubna,  
141980 Dubna, Russia} 
} 
 
\author{Jens Berdermann}{ 
  address={DESY, Platanenallee 6, D-15738 Zeuthen, Germany} 
} 
 
\begin{abstract} 
We present results for neutrino emissivities and bulk viscosities of a 
two-flavor color superconducting quark matter phase with isotropic  
color-spin-locked (iso-CSL) single-flavor pairing which fulfill the  
constraints on quark matter derived from cooling and rotational evolution  
of compact stars. 
We compare with results for the phenomenologically successful, but yet  
heuristic 2SC+X phase. 
\end{abstract} 
 
\maketitle 
 
 
\section{Introduction} 
 
Recently, astrophysical observational programmes monitoring compact stars (CS)  
have provided new, high-quality data for the static properties as well as 
the thermal and spin evolution which put tight constraints on the equation  
of state (EoS) and transport properties of dense matter in CS interiors  
\cite{Klahn:2006ir}. 
In particular, circumstantial evidence for large masses and radii of compact  
stars \cite{Trumper:2003we,Ozel:2006bv} suggests that the EoS at high densities 
must be rather stiff, possibly excluding quark matter interiors. 
In this debate, it has been demonstrated that modern EoS for quark matter  
allow for extended quark matter cores of CS while satisfying mass and radius  
constraints \cite{Alford:2006vz,Klahn:2006iw,Blaschke:2007ri,Grunfeld:2007jt}. 
The resulting stiff hybrid star EoS suffers, however, from the  
{\em masquerade} problem \cite{Alford:2004pf}:  
the corresponding hybrid stars appear to have  
almost identical static properties when compared with pure neutron stars. 
 
A diagnostic tool more sensible than the EoS are the thermal and 
transport properties of dense matter which determine, e.g., the cooling and  
spin evolution of CS. If quark matter occurs in CS interiors we expect it to 
be in a color superconducting state which entails a dramatic dependence on  
the pairing pattern and the sizes of pairing gaps. 
In this contribution, we will focus on the discussion of direct Urca neutrino  
emissivities and bulk viscosities of color superconducting quark matter. 
 
The numerical analysis is based on a Nambu--Jona-Lasinio (NJL) type model, 
allowing a consistent determination of the density and temperature dependence 
of quark masses, pairing gaps and chemical potentials under neutron star  
constraints. The resulting phase diagram suggests that three-flavor phases 
of the CFL-type occur only at rather high densities  
\cite{Ruester:2005jc,Blaschke:2005uj} and render hybrid star  
configurations gravitationally unstable \cite{Buballa:2005,Klahn:2006iw}. 
Moreover, due to large pairing gaps in CFL quark matter, the r-mode  
instabilities cannot be damped  \cite{Madsen:1999ci} and cooling is  
inhibited \cite{Blaschke:1999qx}. 
 
Therefore, we will focus on two-flavor color superconducting 
phases in compact stars, the 2SC+X phase  \cite{Grigorian:2004jq},  
for which a detailed investigation of the cooling phenomenology for  
hybrid stars has already been worked out \cite{Popov:2005xa,Blaschke:2006gd},  
and the iso-CSL phase  \cite{Aguilera:2005tg,Aguilera:2006cj} for which  
a consistent microscopic calculation of the direct Urca emissivity and the  
bulk viscosity will be presented here for the first time  
\cite{Berdermann:2007}.  
This will form the basis of further  phenomenological applications in CS 
physics. 
 
\section{Quark matter in the iso-CSL phase}  
Single-flavor spin-one pairing channels have been investigated before, e.g.,  
in \cite{Schafer:2000tw,Alford:2002rz,Schmitt:2004et}. 
We consider the iso-CSL phase    
\cite{Aguilera:2005tg,Aguilera:2006cj}  
for which in the pairing gap matrix  
$\hat{\Delta}=\Delta(\gamma_3\lambda_2+\gamma_2\lambda_5+\gamma_1\lambda_7)$  
the three antisymmetric color matrices $(\lambda_2, \lambda_5,\lambda_7)$ 
are locked to the three spin matrices ($\gamma_3,\gamma_2,\gamma_1$). 
The two flavor channels are decoupled and the thermodynamical potential is 
therefore   
\begin{equation} 
\label{omega} 
\Omega_q(T,\mu)=\sum\limits_{f=u,d} \Omega(T,\mu_f)~, 
\end{equation} 
where the contribution of a single flavor in mean field approximation  
is given by 
\begin{equation} 
\label{omegaf} 
\Omega(T,\mu_f)=\frac{\bar{\sigma}_f^2}{8G_S}+\frac{3\Delta_f^2}{8G_D} 
-\sum\limits_{r=1}^6\int \frac{d^3~p}{(2\pi)^3}  
\biggl[E_{f,r}(p)+2T{\rm ln}(1+e^{-E_{f,r}(p)/T})\biggr], 
\end{equation} 
The most important feature of the energy spectrum is that the dispersion  
relations of all modes $E_{f,r}$ have nonvanishing gaps and the lowest  
excitation energy is of  ${\mathcal O}$(1 MeV), as required from cooling  
phenomenology. 
Without a microscopic basis, such a mode spectrum has been postulated within  
the 2SC+X phase \cite{Grigorian:2004jq}.  
In the following we compare results for direct Urca neutrino emissivities and  
bulk viscosities of these two phases, using the parameter set for constituent quark mass $M(p=0)=380$ MeV from Ref. \cite{Grigorian:2006qe}. 
 
\subsection{Neutrino emissivity}  
 
Following Iwamotos seminal paper \cite{Iwamoto:1982} where the direct Urca 
emissivity of quark matter,  
\begin{equation} 
\epsilon_0=\frac{457}{630}\alpha_s G_F^2~\mu_e\mu_u\mu_d~ T^6~,  
\end{equation} 
has first been derived, there have been a number 
of calculations, in particular for color superconducting phases. 
From the most recent ones beyond the exponential suppression ansatz, we refer  
to \cite{Jaikumar:2006,Schmitt:2005,Wang:2006}. 
However, none of these is useable for cooling simulations because they  
have either ungapped modes which result in too fast cooling or the pairing  
pattern is not microscopically founded.  
Nevertheless, in deriving the neutrino emissivities for the 2SC+X and the  
iso-CSL phase we follow the strategy of these Refs. with the result 
\begin{eqnarray}\label{emissi}  
\epsilon_{\rm Urca}&=& \epsilon_0~ G_3(\Delta_u,\Delta_d),  
\end{eqnarray}  
where we introduced the function     
\begin{eqnarray}  
G_n(\Delta_u,\Delta_d)&=& 
\frac{5040}{1371\pi^6}\int\limits_0^{\infty}dz~z^n  
\left[\mathcal{F}_1(z)+\mathcal{F}_3(z)+\mathcal{F}_5(z)\right]  
\end{eqnarray}  
with  
\begin{eqnarray}\label{integral1}  
\mathcal{F}_r(z) = \sum\limits_{e_1,e_2=\pm}\int_0^{\infty}\int_0^{\infty}  
dxdy &&(e^{-e_1\sqrt{y^2+a_{u,r}\Delta_u^2}}+1)^{-1}(e^{e_2\sqrt{x^2+a_{d,r}\Delta_d^2}}+1)^{-1}\nonumber\\  
&&\times(e^{z+e_1\sqrt{y^2+a_{u,r}\Delta_u^2}-e_2\sqrt{x^2+a_{d,r}\Delta_d^2}}+1)^{-1},  
\end{eqnarray}  
characterising the influence of the superconducting gaps on the corresponding  
 emissivity.  
For the iso-CSL phase, the coefficients $a_{u,r}$ and $a_{d,r}$ for $r=1,3,5$  
are defined in Ref. \cite{Aguilera:2005tg} and the gaps, obtained from the  
minimization of   (\ref{omega}) fulfill in general    
$\Delta_u \neq \Delta_d$.  
In the 2SC+X phase $\Delta_u=\Delta_d=\Delta$ and $a_{f,1}=a_{f,3}=1$,  
$a_{f,5}=(\Delta_X/\Delta)^2$ for $f=u, d$. 
The density dependent X-gap $\Delta_X$ was introduced in Ref. 
\cite{Grigorian:2004jq} for the first time to appropriately fit the cooling 
data of CS. 
Here we use the parametrization denoted as model IV in Ref. 
\cite{Popov:2005xa}, where $\Delta_X$ has been investigated more detailed to 
fulfill constraints from recent cooling phenomenology.   
The influence of the temperature dependence is taken into account by    
\begin{equation}\label{tdepgap}  
\Delta(T)=\Delta_0\sqrt{1-(T/T_c)^{\beta}},  
\end{equation}  
where one can find values for $\beta$ between 1-3.2 in the literature. In the  
 following calculations we have used $\beta=1$.  
\begin{figure}[hbt]\label{figs1}  
\begin{tabular}{rl}  
  \includegraphics[height=.45\textheight,angle=-90]{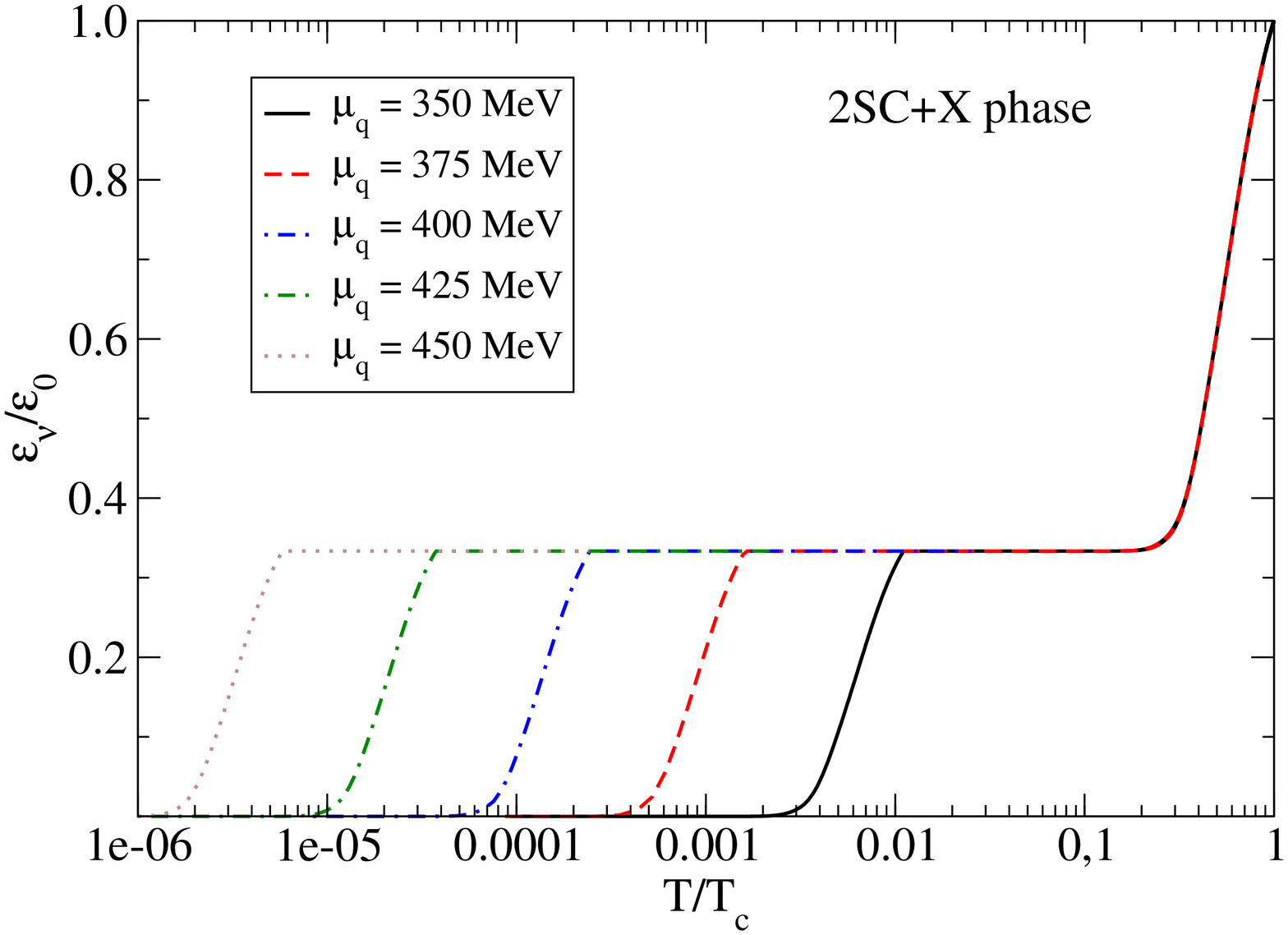}  
\hspace{-1.5cm} 
&\includegraphics[height=.45\textheight,angle=-90]{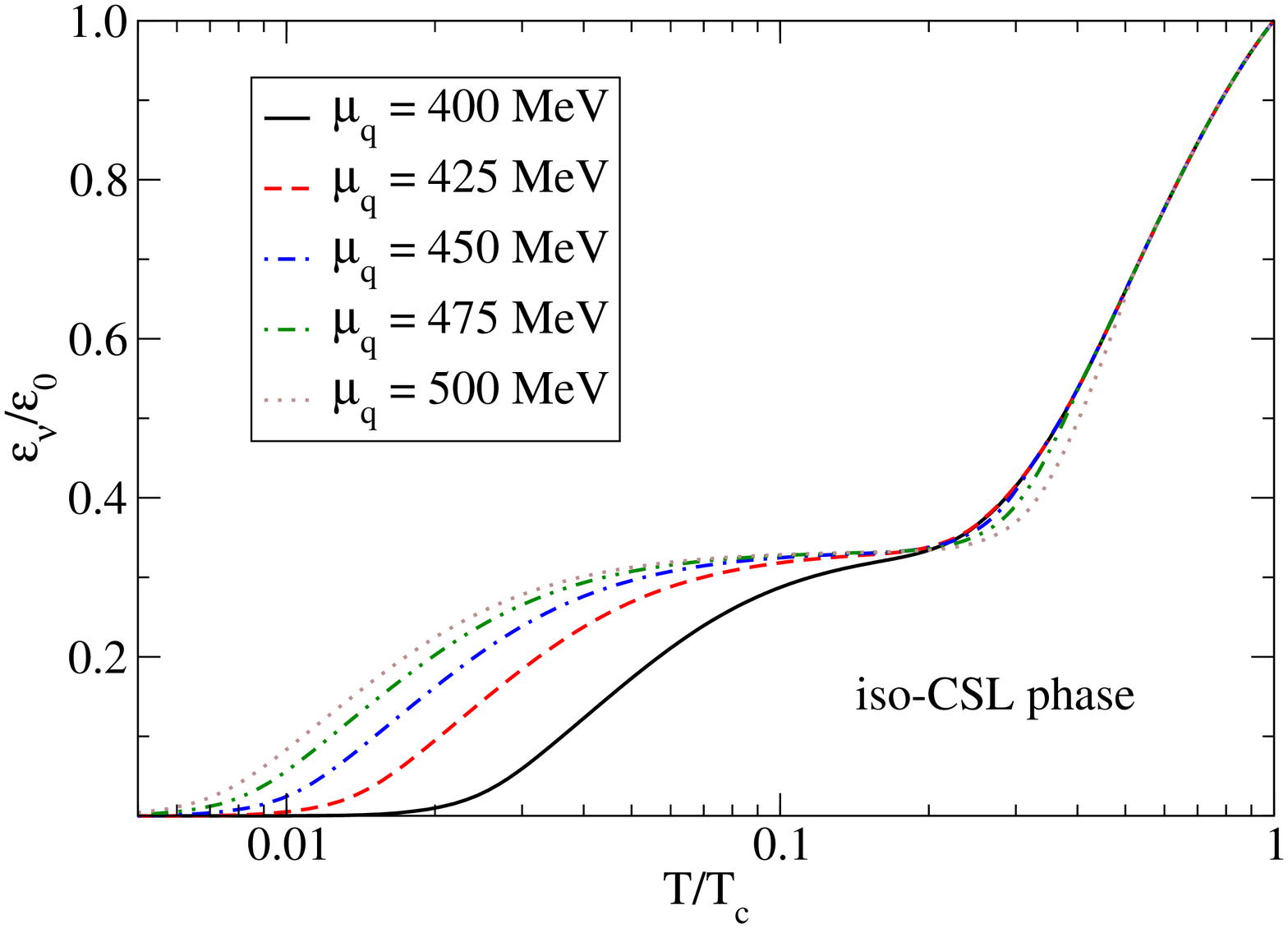}  
\end{tabular}  
 \caption{Neutrino emissivities due to direct Urca processes in the 2SC+X 
phase (left) and in the iso-CSL phase (right).}  
\end{figure}  
In Figure \ref{figs1} we show the emissivities for the microscopic iso-CSL  
phase (right panel) in comparison with the purely phenomenological 2SC+X phase 
(left panel) as a function of temperature for different chemical potentials.  
For both phases a similar suppression of the emissivity is obtained. 
Hence the iso-CSL phase is probably able to explain recent cooling data in a  
more consistent way supporting the idea of superconducting phases in quark  
stars as explanation for observed fast CS cooling.    
  
\subsection{Bulk viscosity}  
  
According to \cite{Andersson:1997xt}, in the absence of viscosity all rotating  
CS would become unstable against r-modes  \cite{Andersson:2000mf}.  
Therefore, from the observation of millisecond pulsars, one can derive  
constraints for the composition of CS interiors  
\cite{Madsen:1999ci,Drago:2007iy}.   
For such an investigation, the bulk viscosity is a key quantity and we want to 
consider it for the two-flavor color superconducting phases introduced above, 
following the approach described in Ref. \cite{Sa'd:2006qv}. 
Note that the 2SC phase considered in   \cite{Alford:2006gy} and in the  
contribution to this volume \cite{Alford:2007pj} is a three-flavor phase,  
where the nonleptonic process $u+d \leftrightarrow u+s$ provides the dominant  
contribution. Due to absence of strange quarks in the 2SC phase of the  
present paper, this process does not occur.  
 
\begin{figure}[hbt]  
\begin{tabular}{rl}  
  \includegraphics[height=.45\textheight,angle=-90]{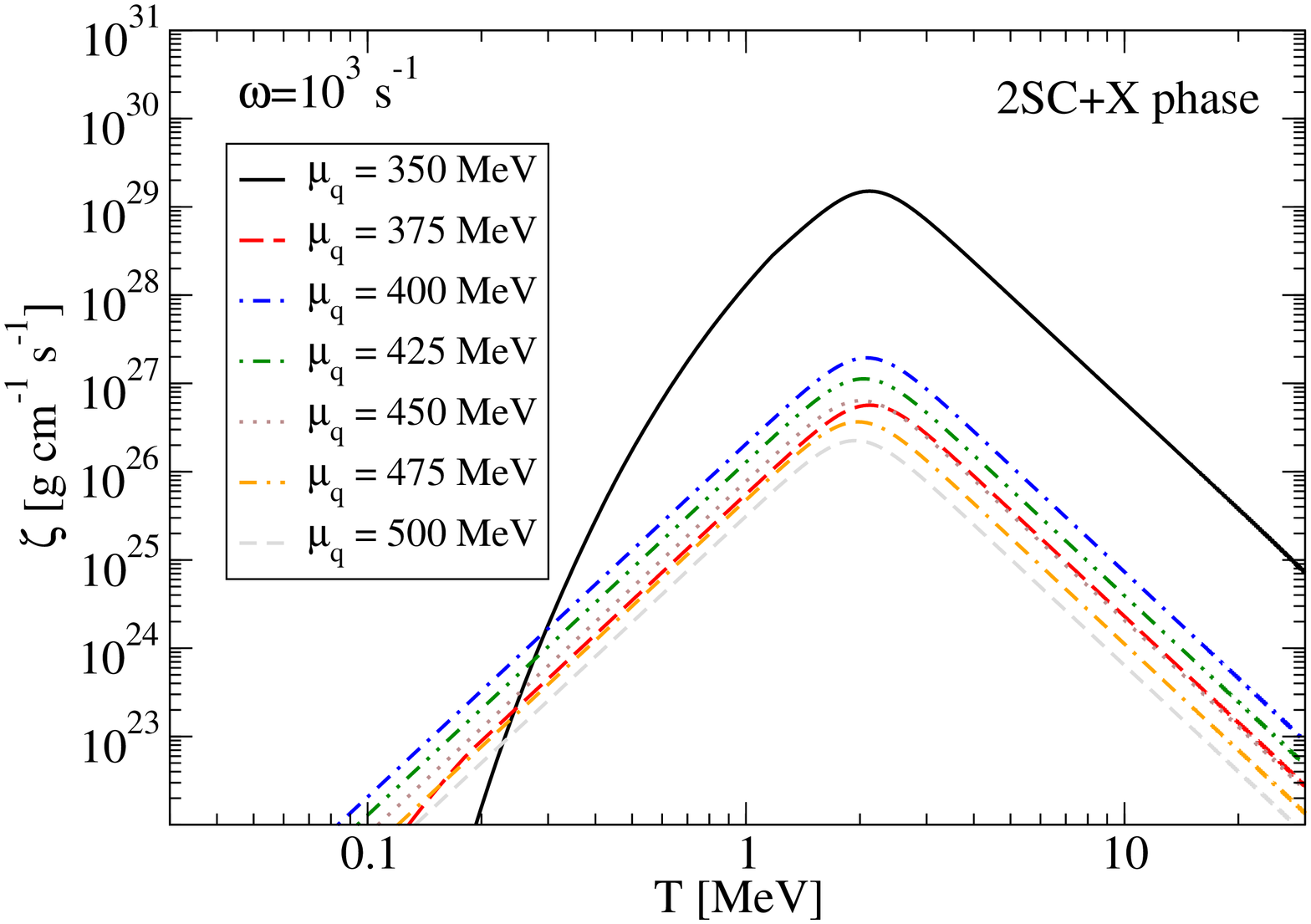}  
\hspace{-1.5cm} 
&\includegraphics[height=.45\textheight,angle=-90]{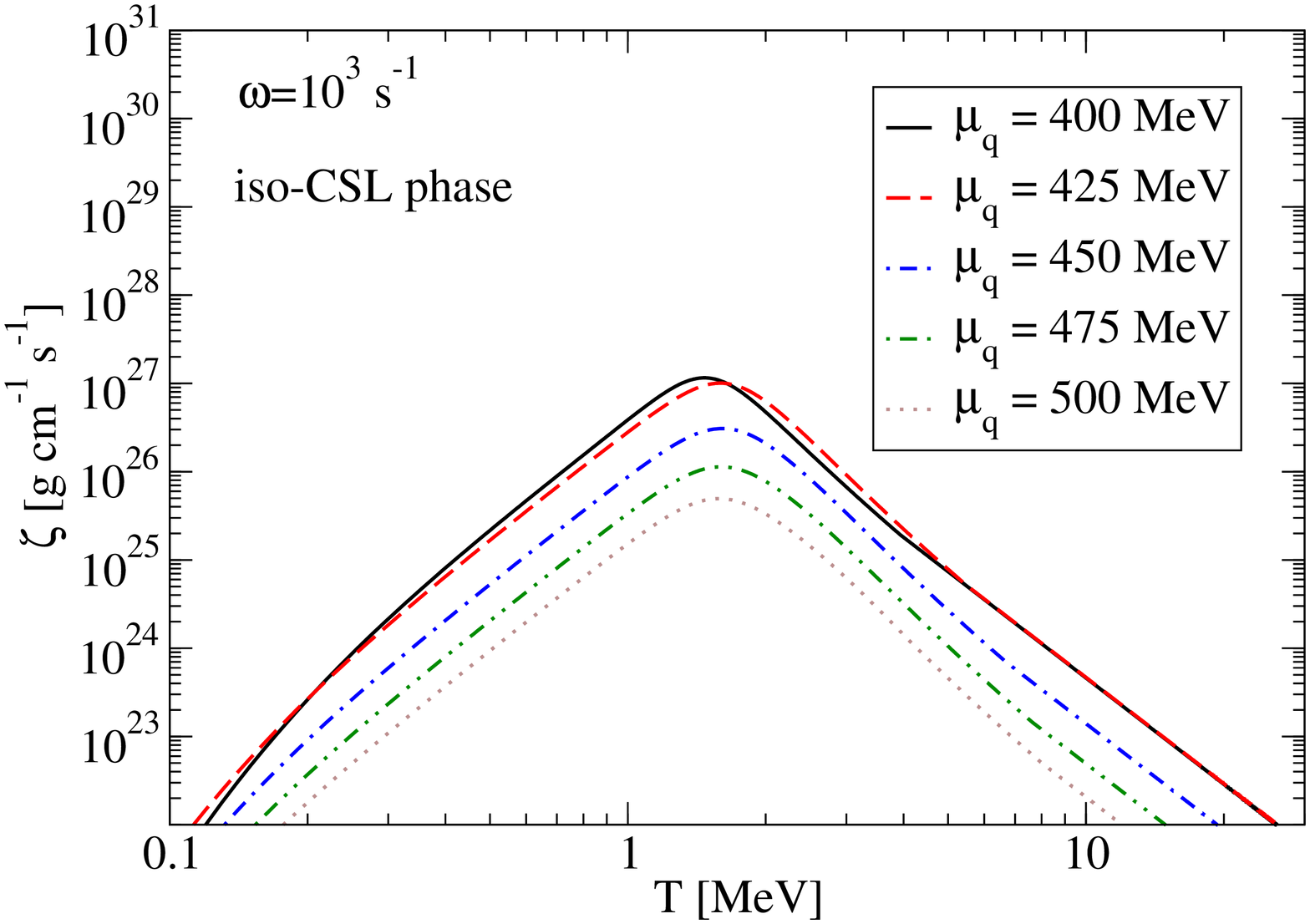}  
\end{tabular}  
 \caption{Temperature dependence of bulk viscosities in the 2SC+X phase   
(left) and in the iso-CSL phase (right) for a frequency of $\omega=1$ kHz,  
typical for excitations of r-modes in millisecond pulsars.}  
\label{viscosity} 
\end{figure}

The bulk viscosity at all temperatures is determined by  
\begin{equation}  
\zeta=\frac{\lambda C_t^2}{\omega^2+(\lambda B/n)^2}  
\end{equation}  
with $C_t=C+C`$ and the coefficients functions  
\begin{eqnarray}  
C&=&\frac{M_u^2}{3\mu_u}-\frac{M_d^2}{3\mu_d},\nonumber\\  
C`&=&\frac{4\alpha_s}{3\pi}\left[\frac{M_d^2}{\mu_d}\left({\rm ln}\frac{2\mu_d}{M_d}-\frac{2}{3}\right)-\frac{M_u^2}{\mu_u}\left({\rm ln}\frac{2\mu_u}{M_u}-\frac{2}{3}\right)\right]\nonumber\\  
B&\simeq&\frac{\pi^2}{3}n\left(\frac{1}{\mu_u^2}+\frac{1}{\mu_d^2}+\frac{1}{\mu_e^2}\right).  
\end{eqnarray}   
The relevant processes for the bulk viscosity in two-flavor quark matter are  
the flavor changing weak processes of electron capture and beta decay, with a 
direct relation to the direct Urca emissivity 
\begin{eqnarray}  
\lambda&=&\frac{3}{2}\frac{\epsilon_0}{T^2}~ G_1(\Delta_u,\Delta_d)~,  
\end{eqnarray}  
The numerical results for the NJL model in selfconsistent meanfield  
approximation are displayed in Fig. \ref{viscosity} for the 2SC+X phase   
(left panel) and the iso-CSL phase (right panel) in striking similarity. 
Note that in comparison with Ref. \cite{Sa'd:2006qv} the  peak  
value of the viscosity is also located at $T= 1 \sim 2$ MeV, but up to three 
orders of magnitude higher! Since the normal quark matter results coincide, 
this must be a result of the selfconsistent treatment of masses, gaps and  
composition (chemical potentials) in the present models. 
In particular the strongly density dependent X-gap is rapidly decreasing with 
increasing density as one can see by the dramatic change for the bulk 
viscosity at low quark chemical potentials. 
 
\section{Conclusion} 
Transport properties in dense quark matter depend 
sensitively on the color superconductivity pairing patterns and provide thus 
a tool for unmasking the CS interiors by their cooling and rotational  
evolution characteristics. 
On the example of neutrino emissivities and bulk viscosities for the 2SC+X and
the iso-CSL phase we have demonstrated that both two-flavor color  
superconducting phases fulfill constraints from the CS phenomenology.   
For the 2SC+X phase with yet heuristic assumptions for the X-gap the hybrid 
star configurations and their cooling evolution have been numerically 
evaluated in accordance with observational data. 
The temperature and density behavior of the neutrino emissivity in the  
microscopically well-founded iso-CSL phase appear rather similar so that we  
expect a good agreement with CS cooling data too. 
The bulk viscosities for both phases have been presented here for the first  
time and provide sufficient damping of r-mode instabilities to comply with 
the phenomenology of rapidly spinning CS. 
We conclude that the subtle interplay between suppression of the direct Urca  
cooling process on the one hand and sufficiently large bulk viscosity puts  
severe constraints on microscopic approaches to quark matter in compact stars.
 
 
\begin{theacknowledgments} 
D.B. likes to thank M.~Alford, R. Anglani, A.~Drago, B.~A.~Sa'd and  
H.~Malekzadeh for 
discussions and is grateful to the organizers for the invitation and for  
the stimulating atmosphere of the meeting.  
He received support from the Polish Ministry for Science and Higher Education. 
\end{theacknowledgments} 
 

\end{document}